\documentclass[aps,showpacs, superscriptaddress]{revtex4}
\usepackage{graphicx}

\begin{document}

\title{Structural and magnetic properties of single dopants of Mn and Fe for Si-based spintronic materials}

\author{M. Shaughnessy}
\affiliation{Department of Physics, University of California, Davis, CA 95616-8677}
\affiliation{Condensed Matter and Materials Division, Lawrence Livermore National Laboratory, Livermore, CA 94551}
\author{C.Y. Fong}
\affiliation{Department of Physics, University of California, Davis, CA 95616-8677}
\author{Ryan Snow}
\affiliation{Department of Physics, University of California, Davis, CA 95616-8677}
\author{L.H. Yang}
\affiliation{Condensed Matter and Materials Division, Lawrence Livermore National Laboratory, Livermore, CA 94551}
\author{X.S. Chen}
\affiliation{Institute of Technical Physics, Shanghai, China}
\author{Z.M. Jiang}
\affiliation{Surface Physics Laboratory, Fudan University, Shanghai, China}

\begin{abstract}
Single dopings of Mn and Fe in Si are investigated using 8-, 64-, and 216-atom supercells and a first-principles method based on density functional theory. Between the two transition metal elements (TMEs), atom sizes play an essential role in determining the contraction or the expansion of neighboring atoms around the TME dopant at a substitutional site. At a tetrahedral interstitial site, there is only expansion. Magnetic moments/TME at the two sites are calculated. Physical origins for these inter-related properties are discussed. A few suggestions about the growth of these Si-based alloys are given.


\end{abstract}

\maketitle

\newpage


\section{Introduction} Since the reports\cite{Park} of epitaxially grown Mn$_x$Ge$_{1-x}$ with $x$ $<$ 3.5\% in crystalline form showing the Curie temperature, T$_C$, up to 116 K ,  Mn$_x$A$_{1-x}$ alloys, where A is a group-IV element, have attracted much attention as potential candidates for spintronic materials.  In particular, there were interesting experiments \cite{Liu, Bolduc, Ma} measuring the transport and magnetic properties of Mn$_x$Si$_{1-x}$ grown by post thermal treatment of amorphous Mn-Si films, ion implantation, and arc-melting methods. In addition, the growth of Fe in Si has also been carried out by Su et al. \cite{Su} who used the molecular beam epitaxial method to successfully grow film forms of Fe$_x$Si$_{1-x}$ with x at 4.0 and 7.0\%, respectively. Their samples show uniform distribution of the transition metal element (TME). Theoretically, there have been many model calculations of Mn doped in Si\cite{Dalpian, da Silva, Bernardini, Weng, ZZhang}. Many of them examined the relative energetics involving more than one Mn atom. Weng and Dong \cite{Weng} also report diverse magnetic properties, such as both ferromagnetic and antiferromagnetic phases, based on energetic arguments, and magnetic moments in Si doped with Cr and Fe. 

Because Si technologies are the most mature among all the semiconductors the prospect of realizing spintronic devices using Si-based alloys is more promising than doping TME in other semiconductors. 
To use a material for fabricating spintronic devices, one should consider the following issues: (a) Can the material be easily grown? (b) How large can the saturation magnetic moment/vol or the magnetic moment/TME be? (c) How can possible distortions near the doping sites, contraction or expansion, be predicted before growth? And, (d) what is the relative stability between the doping sites, in particular a substitutional (S) and a tetrahedral interstitial (I) sites? There is also a higher energy and less stable six-fold coordinated interstitial site \cite{Leung} which we do not consider in this paper. 

Why is (c) important? Our earlier work on MnC\cite{Qian} shows that small volume around Mn can diminish the magnetic moment of the compound because there is not enough space available for all electrons localized at the TME to align their spins, causing spin flips and reducing the saturation magnetic moment. This also demonstrates that issues (b) and (c) can be inter-related. Such entangled issues make it difficult to use only physical intuition to guide the growth. Furthermore, there is a calculation indicating that an S site is less stable than the I site based on energetic considerations\cite{Wu}. However, we found that in order to explain the large magnetic moment/Mn determined by Bolduc et al. \cite{Bolduc} in Mn$_x$Si$_{1-x}$ with x = 0.1\% it is necessary for the Mn to occupy an S site\cite{Shaughnessy}. Despite such complications, exploring more new spintronic materials by doping various TMEs in Si is still an appealing option.

Because the issues involved are just coming into focus, it would be a major undertaking to use a trial and error approach to grow new Si-based alloys without a judicious choice of TMEs. In this paper, we provide a basic understanding of the energetic and magnetic properties of singly doped Fe and Mn in Si.  A brief discussion of the models and the method of calculation are given in Section II. Results and discussion will be given in Section III. In Section IV, we summarize the results and make some suggestions concerning the growth.
	\section{Models and method of calculation}
	We construct models of single TME dopants by starting with a conventional cell of the diamond crystal having 8-atoms/unit-cell and increasing the cell size up to 216-atom/unit-cell by stacking the conventional cells in three directions. As an example, models for the 8-atom case including the atoms at the boundaries of the unit cell and with one TME at the S and I sites are shown in Fig. 1 on the left and the right, respectively. The different supercell sizes simulate in some way the degree of isolation of the single atom dopants in the alloys, although the supercell approach precludes directly modeling random alloys because it is implicitly periodic ordered in real space. There is a possibility that disorder in the real alloys may affect the validity of our results, particular at high concentrations. The popular coherent potential (CPA) \cite{Gyorffy} and virtual crystal approximations \cite{Bellaiche}, include disorder through scattering or crystal structure, but both of these may introduce unphysical effects. A more direct method is the special quasirandom structures (SQS) approach \cite{Zunger}. Previous work \cite{Chroneos1} found deviations of order 0.01\space \AA \space from Vegard's Law, which states that the alloy lattice constant scales linearly with $x$ between the A and B lattice constants for a binary alloy of $A_xB_{1-x}$, in group IV semiconductors alloyed with Sn using the SQS method. We expect that including disorder may change our results by less than 0.01\space \AA, because some deviation from Vegard's Law can already be accommodated within the ordered alloys we model. In any case, our single dopant results will be most reliable for low concentration limits, in which the random nature of the alloys should be unimportant.  	\begin{figure}
  \includegraphics[angle=0,width=3in,clip=true]{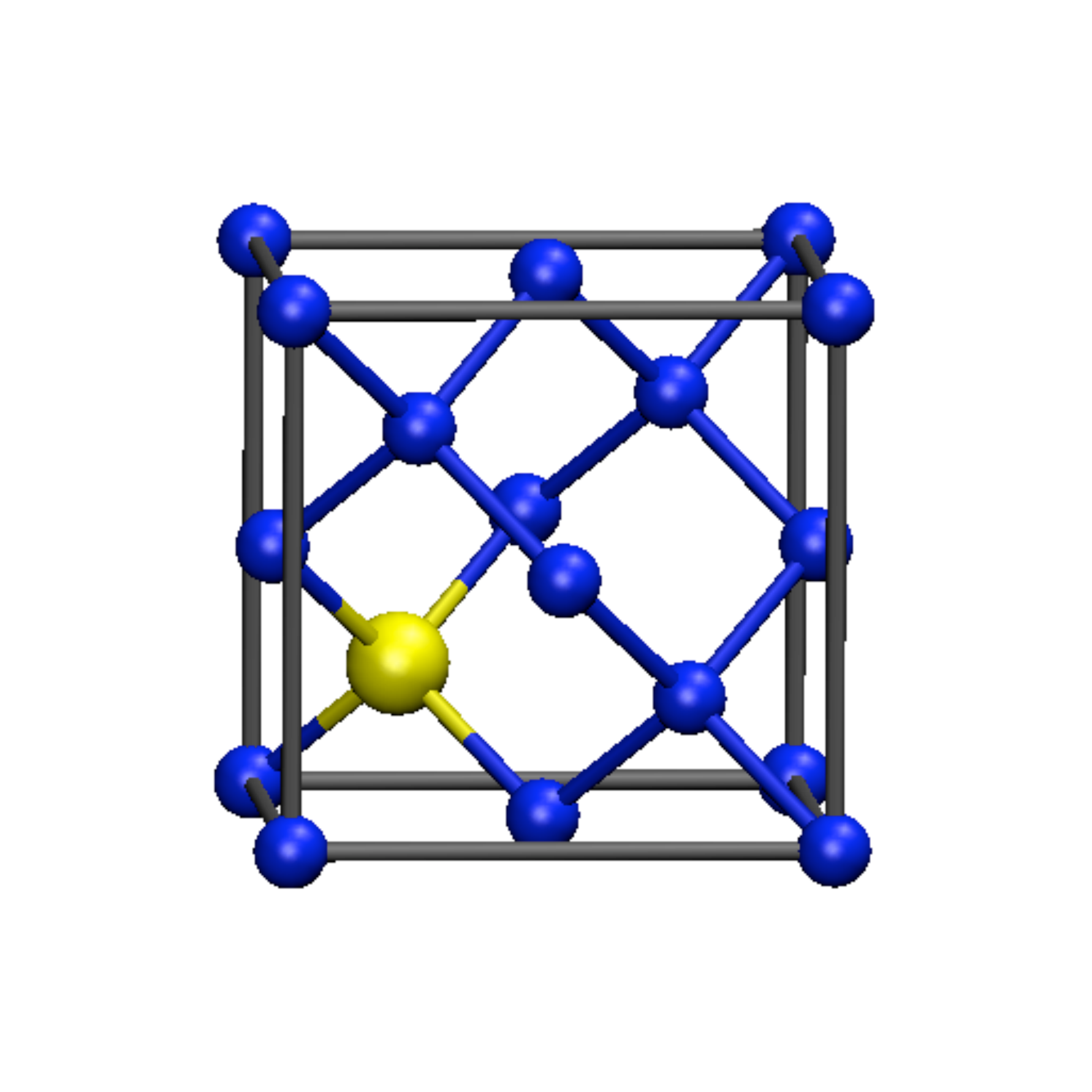}
    \includegraphics[angle=0,width=3in,clip=true]{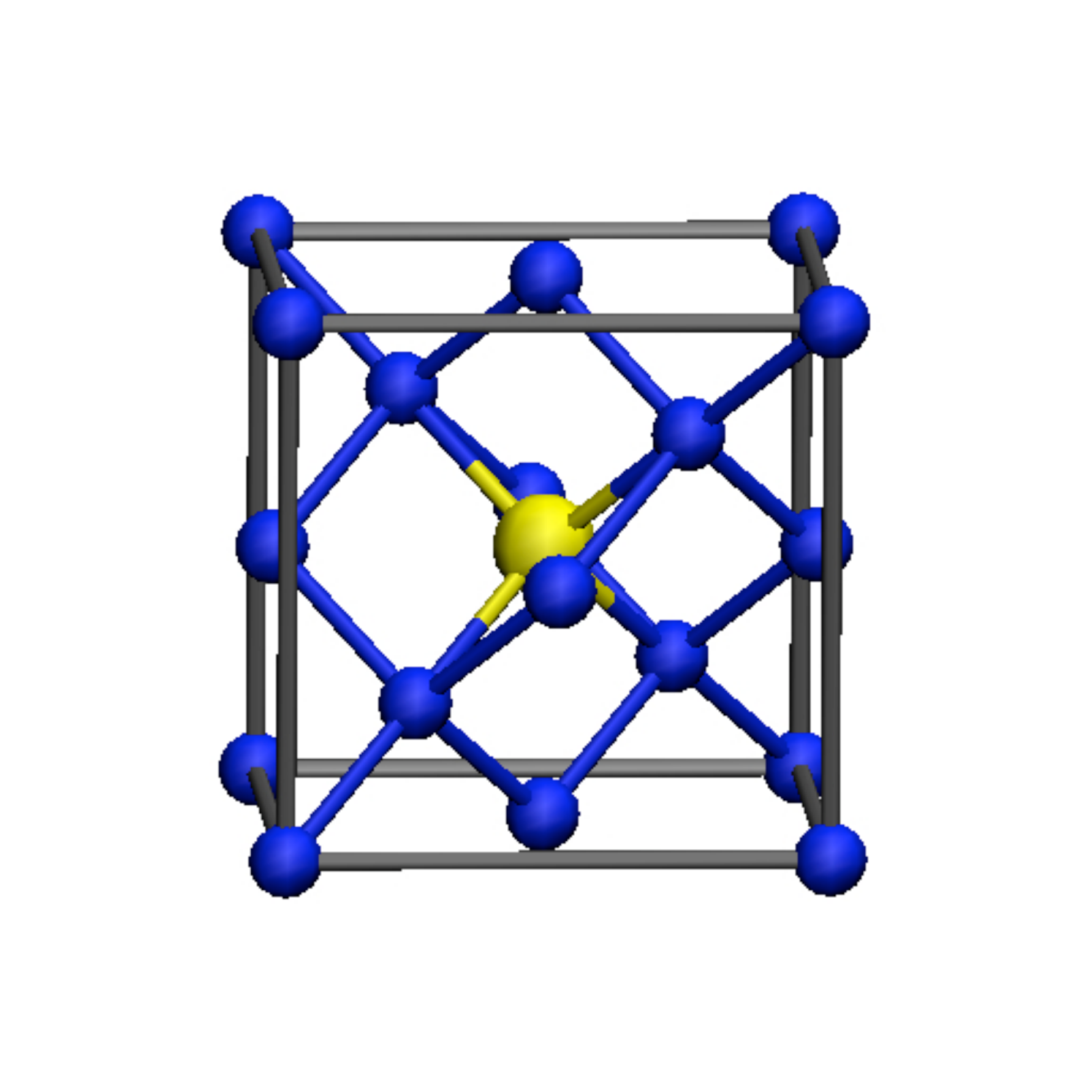}
  \caption{Left: 8-atom cell with TME at an S site. The TME is the light (yellow online) sphere and the Si atoms are the darker (blue online) spheres. Right: 8-atom cell with TME at an I site.}
 \label{fig:1}
\end{figure}

	We used the VASP code \cite{VASP} to determine the basic structural and magnetic properties of the alloys using ultrasoft pseudopotentials of Si, Fe and Mn with the normal electronic configurations. The generalized gradient approximation (GGA) of Perdew 91\cite{Perdew} was used to treat the exchange-correlation of the electrons. A planewave basis with 650 eV cutoff energy and a (15,15,15) Monkhorst-Pack \textbf{k}-point mesh \cite{Monkhorst} was used for the 8-atom case. As the cell size increased, we reduced the mesh points according to the inverse ratio of the lattice constant of the large cell to the one of the 8-atom case. Because of the doping, the forces acting on the atoms may initially be large. We therefore relax the positions of the atoms in each unit cell to reduce the forces acting on all atoms to less than 6.0 meV/\AA.
	
	In the spirit of first-principles calculations, the lattice constant of the crystalline Si was optimized with respect to the total energy. It is 5.45\AA, which is 0.4\% larger than the experimental value. This lattice constant was used to determine the bond length, 2.36\AA, between any pair of Si atoms in the supercells before the TME was inserted and the force relaxation was subsequently carried out. 
		\section{Results and discussion}
	\begin{table*}
\caption{Relaxations and bond lengths}
\begin{center}
\begin{tabular}{cccccc}
\hline
\hline
TME &Size&Position \space \space& Bond length(\AA) \space \space& Relaxation (\AA) \space \space& Lattice constant change(\%)\\
\hline
Mn & 8 & S &2.38& 0.02& 0.58  \\
 &  & I &2.41& 0.05& 0.95  \\
& 64 & S &2.40& 0.04& 0.17  \\
 &  & I &2.40& 0.04& 0.16  \\
 & 216 & S &2.40& 0.04& 0.08  \\
&  & I &2.43& 0.07& 0.13  \\
Fe & 8 & S &2.32& -0.04& -0.69  \\
 &  & I &2.36& 0.00& -0.70  \\
& 64 & S &2.26& -0.10& -0.05  \\
 &  & I &2.40& 0.04& 0.11  \\
 & 216 & S &2.25& -0.11& 0.06  \\
&  & I &2.40& 0.04& 0.11  \\
\hline
\hline
\end{tabular}
\end{center}
\end{table*}
	\subsection{Comparison of substitutional and interstitial sites}

For each supercell, we considered two doping sites for the TME: the S and I sites. Table I contains the bond length between a TME and its nearest neighbor (nn) Si and the change from the pure Si value after the relaxations of atomic positions. We also independently optimized the lattice constant for each model. We show later that it enables us to understand the experimental result in \cite{Su} because the experiment measures the change in lattice constant under alloying. This data is shown in the last column. Except for the I site of the 8-atom case and the S site of the 216-atom case for the Fe doping, the qualitative features are consistent with the local bond length changes. The change in lattice constant in the 8-atom cases is much larger than in either the 64- or 216-atoms cases, suggesting that for $x \gtrsim 12\%$ interaction between dopants can be significant. The relatively small supercell size for these 8-atom cases produces large strain fields on the supercell boundary. We now discuss the differences between the two doping sites and the two dopant species.

				\begin{figure*}
  \includegraphics[angle=0,width=6in,clip=true]{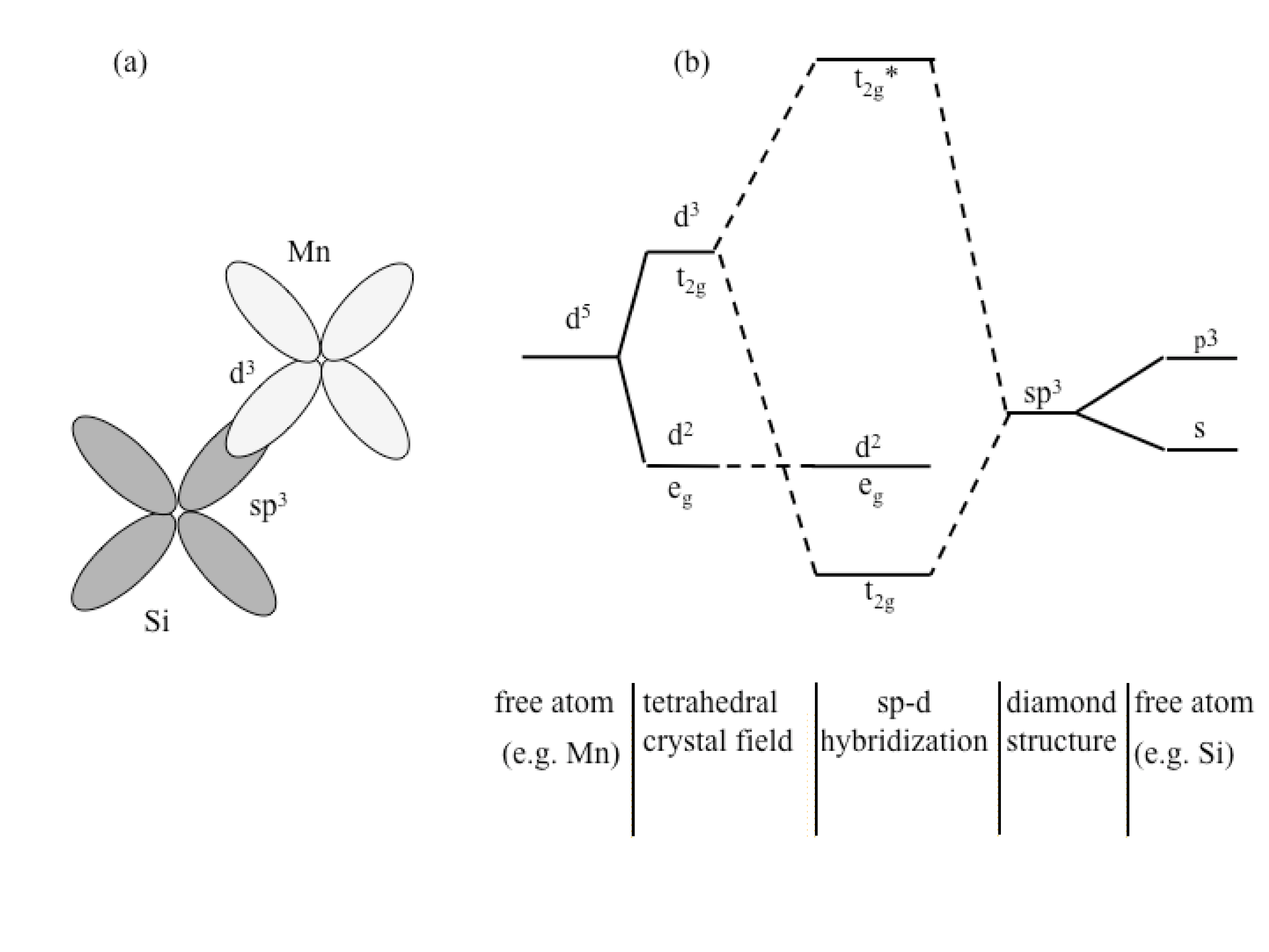}
  \caption{(a) Diagram showing the d-p hybridization between the Si atom at (0.0,0.0,0.0)a and the TME at (1/4,1/4,1/4)a, where a is the lattice constant of the conventional cell. (b) Schematic diagram showing the crystal field and hybridization effects.}
 \label{fig:3}
\end{figure*}
	At both the S and I sites, four Si atoms surround a TME, providing a tetrahedrally symmetric environment. The tetrahedral crystal field effect causes the five-fold degenerate d-states of the TME to split into two-fold, e$_g$, and three-fold, t$_{2g}$, states for both sites. The interactions, however, between the TME and its nn Si atoms differ dramatically at the two sites. Only at the S site do the t$_{2g}$ states combine with the s-states of the TME hybridize with the sp$^3$ orbitals of the nearest nn Si atoms to form the bonding and antibonding states. A simple picture of the hybridization is shown in Fig. 2(a). The combined effect of the crystal field and the hybridization on the energy levels is schematically given in Fig. 2(b). For the I site on the other hand, a TME cannot make strong d-p hybridized bonds with its nn Si atoms because each nn Si has already saturated its bonds with its own Si neighbors. 	
	
					\begin{figure}
  \includegraphics[angle=0,width=4in,clip=true]{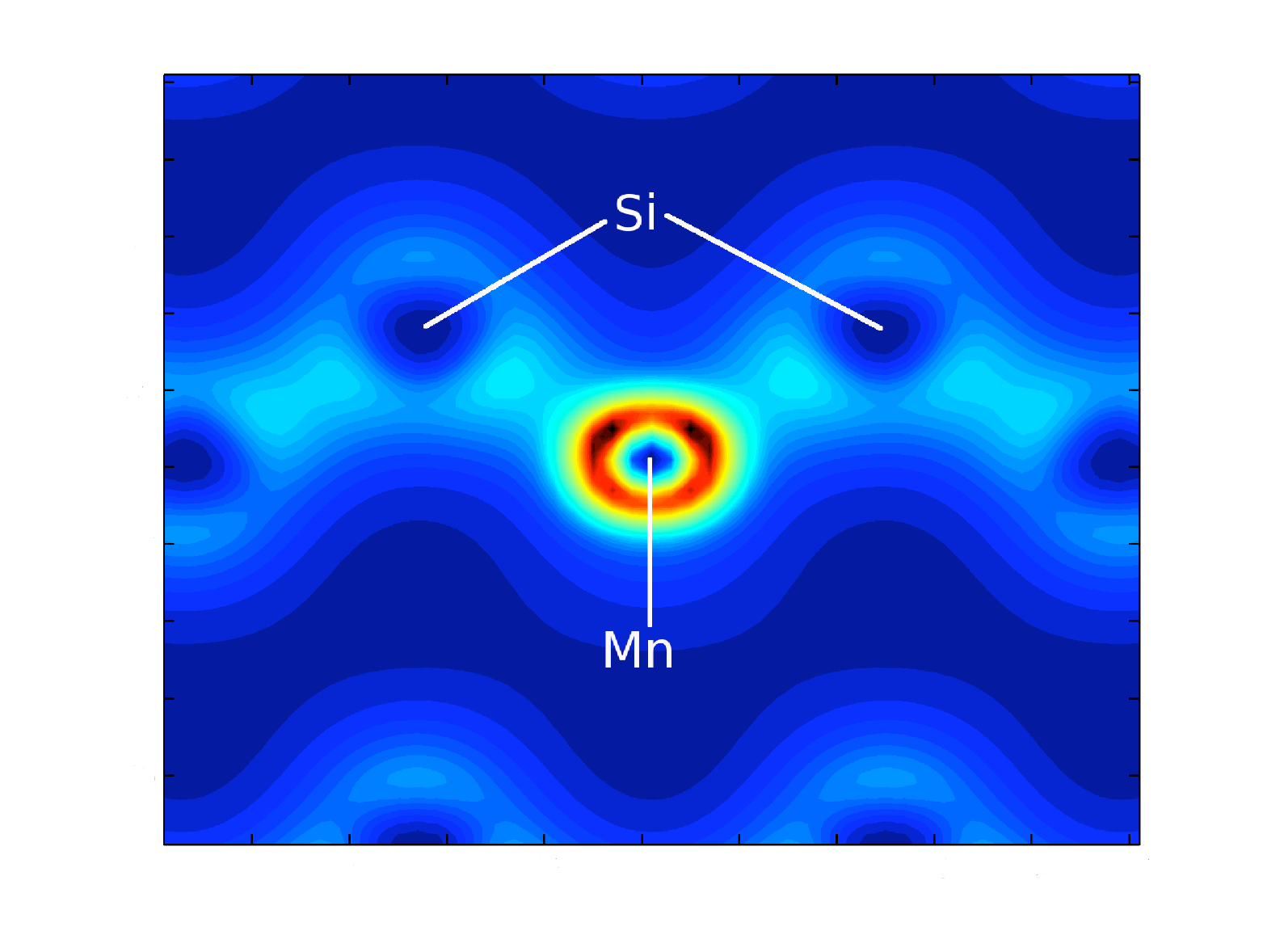}
  \caption{Total charge density of Mn at an S site in the 64-atom model. The section is formed by [110] and [001] axes containing Mn at an S site.}
 \label{fig:4}
\end{figure}
				\begin{figure}
  \includegraphics[angle=0,width=4in,clip=true]{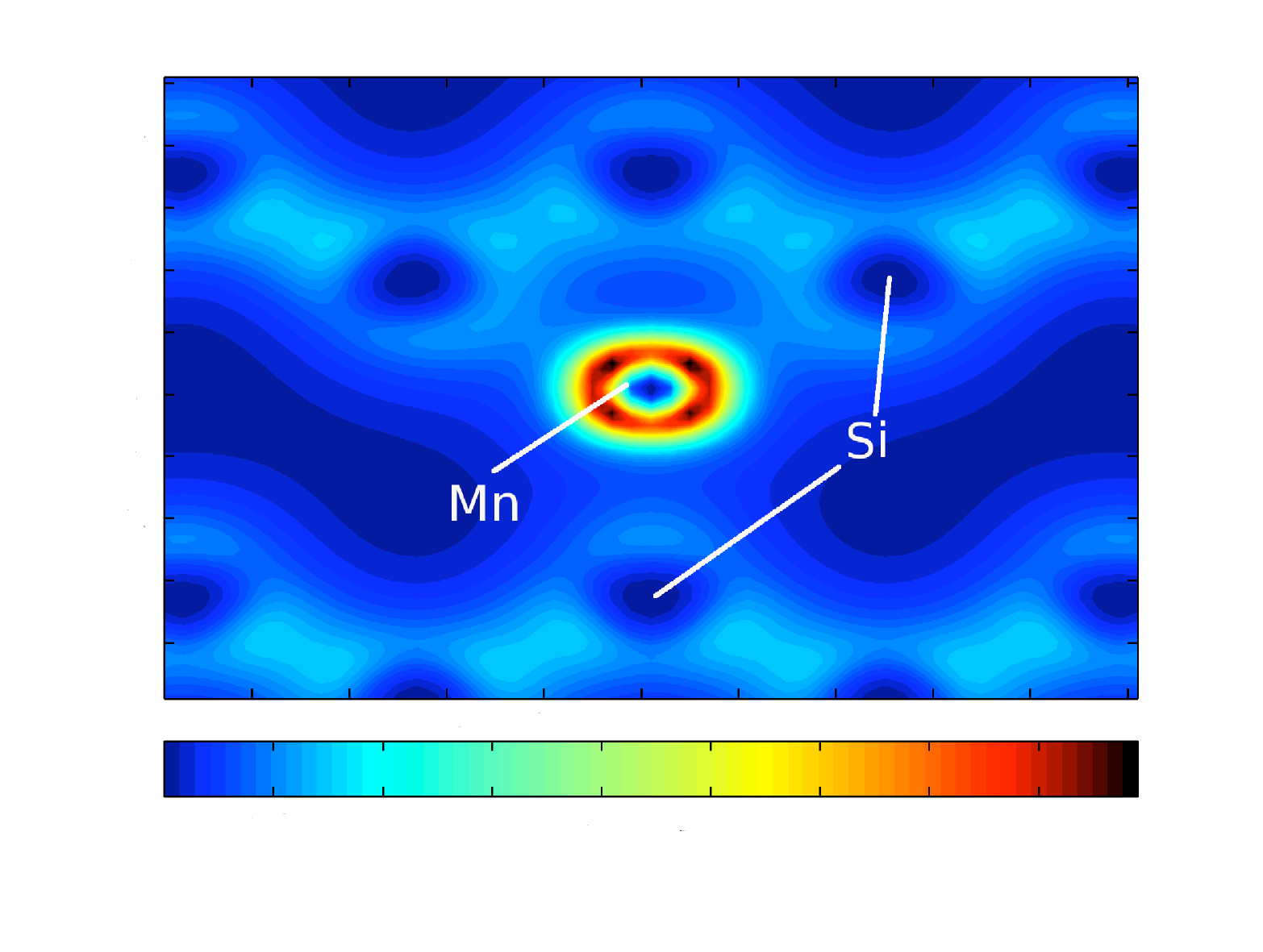}
  \caption{Total charge density of Mn at an I site in the 64-atom model. The section is formed by [110] and [001] axes containing the Mn at an I site. The relative coloring scale for coloring the contours is shown below the plot.}
 \label{fig:5}
\end{figure}
	
	The effects of the environment on the Mn dopant are shown by the total charge distribution in Figs. 3 and 4. The qualitative features for the Fe are similar. In Fig. 3, the charge distribution of Mn doped at an S site in the 64-atom model is plotted in the plane formed by the [110] and the [001] axes of the supercell. The Mn and some of the Si atoms are indicated. The two dark spots close to the Mn atom are maxima of the charge density coming from the d$_{xz}$- and d$_{yz}$-type orbitals.
	
	Two of these lobes point toward the two in-plane nn Si. The bond charge from the d-p hybridization is close to the Mn as indicated by the dark spots. The sp$^3$ orbitals of the nn Si atoms participating in the d-p hybridization appear in the form of light regions between the two in-plane nn Si atoms and the Mn. The bond charge between any other two neighboring Si atoms forms a distinct light-colored dumbbell shaped region characteristic of the sp$^3$ orbitals associated with the Si atoms. In Fig. 4 the total charge distribution of Mn at an I site is given in the same section as in Fig. 3. Again the density near the Mn shows the dense contours and four dark lobes characteristic of the d-states. The relevant feature is in the region between the Mn and the four nearby Si atoms. Here there is no indication of strong hybridization between the Mn and the Si. Rather the valence electrons are spread out into the interstitial regions under the attractive interaction from the Si nuclei due to the relatively open diamond structure. These electrons interact with the nearby bond charges to produce the expansions in Table I.  
	
	The distinct bonding features of the S and I sites are the physical basis underlying the fact that the I site is favored during the growth of these alloys. In order for a TME to occupy an S site, it is necessary to break the d-p hybridized bonds. At an I site, no such breaking process is required. The energy required to expand the nn distance between the TME and the Si atoms is definitely smaller than the one to break bonds.

	\subsection{Comparison between doping Mn and Fe}
	  
In Table I the differing bond length relaxation for the Mn and Fe dopants is evident. The distance to the nn Si atoms always increases with Mn at either site, but decreases when the Fe is at an S site. The lattice constant change data in the last column is consistent, showing that the supercells with Mn always expand in volume. For Fe doping, the supercell volume change depends on the size of the supercell and the site of the Fe atom. For the 216-atom supercell, both doping sites show expansion. The contraction of the lattice constant for the smaller supercells indicates that the neighboring unit cells interact, suggesting alloys with larger concentrations may contract in volume when Fe is doped. 

Remembering that the optimized bond length for pure Si is 2.36 \AA, a general feature is the bond length increase/volume expansion with Mn at an S site. The small size of the 8-atom model constrains the expansion more than in the 64- and 216-atom cases. For doping Fe at an S site, the bond length, in general, decreases. The size of the supercell determines the magnitude of the supercell volume contraction. In the larger 64- and 216-atom models, corresponding to x = 1.6\% and 0.46\% the volume expansion is small indicating the relaxation is not constricted. At this point, we suggest that to model an alloy using a supercell, the size of the supercell should be consistent with the concentration specified in experiment. Comparing to standard atomic radii \cite{PTable}, 1.26 \AA \space for Fe, 1.35 \AA \space for Mn and 1.32 \AA \space for Si, the behaviors of the calculated bond lengths between the S site TMEs and their nn Si atoms can be understood. In particular, we note that this bond length increases when the TME has a larger atomic radius (Mn) and decreases when the atomic radius is smaller (Fe).

	Since the calculated relaxed bond lengths at an S site and an I site for Fe show opposite character regardless of supercell size, it can be used to identify the sites of the Fe doped in Si. In \cite{Su} with 4.0\% doping of Fe, the lattice constant expends by 0.3\%.  We suggest that majority of the Fe in the alloy occupies interstitial sites. Also, in \cite{Ma}, the authors suggested that the Mn occupies S sites because the measured lattice constant of their alloy is larger than the Si lattice constant. As we show, the lattice constant of the alloy will expand regardless of whether the Mn atom occupies either an S or an I site.
	

	\subsection{Energetics}
	
		To form an alloy, the primary energetic concern is the formation energy difference of doping Fe and Mn at the two different sites, respectively. We simplify the chemical potential type of approach \cite{GQian} and use only the chemical potential of Si. In Table II, the calculated cohesive energies of the Mn and Fe in different size superlattices and at different sites are presented. It is reassuring to note that when we subtracted the energies of two different supercell sizes with the TMEs either at both S or both I sites and divided by the difference of the number of atoms, we obtained the energy/Si. The largest deviation from the -5.433 eV/Si of the crystalline Si is 0.07\%. The -5.433 eV/Si is thus used as the chemical potential of Si, $\mu_{Si}$. The formation energies of the I site or the S site are given by: 
		\begin{equation}
E_{f} +\mu_{TME}  = E_{coh} - N\mu_{Si},	
\end{equation}
where $E_{f}$ is the formation energy of an alloy, $E_{coh}$ is the cohesive energy of an alloy, and N is the number of Si atoms in the supercell. The formation energy difference between the S and I models is then
\begin{equation}
\Delta E^{I-S}_{f} = E_{coh}^I - (E^S_{coh} + \mu_{Si}).
\end{equation} 

Table II also summarizes the difference of formation energies at different sites for the three supercells. The I site consistently has lower energy than the S site. The difference is, in general, about 0.500 eV except the value of 0.164 eV for the 8-atom Mn case. The constricted expansion in the small 8-atom cell causes the smaller difference.

	 \begin{table*}
\caption{Formation energy differences and magnetic moments (M) after supercell relaxation}
\begin{center}
\begin{tabular}{ccccc}
\hline
\hline
TME &Size&Position\space \space \space& E$_f$(I)-E$_f$(S) (eV)& M ($\mu_B$/TME)\\
\hline
Mn & 8 & S & -0.164&2.97  \\
 &  & I &&3.16   \\
& 64 & S & -0.459&3.00  \\
 &  & I &&3.00 \\
 & 216 & S &-0.528&3.00  \\
&  & I &&3.00   \\
Fe & 8 & S & -0.415&1.76  \\
 &  & I &&2.05  \\
& 64 & S & -0.503&0.00  \\
 &  & I &&2.00  \\
 & 216 & S & -0.473&0.00  \\
&  & I &&2.00   \\
\hline
\hline
\end{tabular}
\end{center}
\end{table*}
	
	\subsection{Magnetic moment}
		
	 \begin{table}
\caption{Experimentally measured magnetic moments of Mn$_x$Si$_{1-x}$}
\begin{center}
\begin{tabular}{ccc}
\hline
\hline
x &Magnetic moment($\mu_B$/Mn)&Reference\\
\hline
0.1\% & 5.00 & \cite{Bolduc} \\
 0.8\%& 1.50 & \cite{Bolduc}   \\
1.0\%& 4.15 & \cite{Ma}  \\
 1.5\%& 4.05& \cite{Ma} \\
\hline
\hline
\end{tabular}
\end{center}
\end{table}

	In general, for the TME at an S site, the ionic model \cite{Fong} applies. Each of the four neighboring Si take one electron away to participate in d-p hybridization leaving the rest of electrons near the TME to align their spins according to the first Hund's rule and reduce their Coulomb repulsion. 
The model predicts the moments for Mn and Fe to be 3.00 and 4.00 ($\mu_B$/TME), respectively. These two results serve as references.

	Refs. \cite{Bolduc, Ma} give the measured magnetic moment/Mn, M, in Mn$_x$Si$_{1-x}$. The values range from the maximum 5.0 $\mu_B$/Mn at x = 0.1\% to 1.5 $\mu_B$/Mn at x = 0.8\%. They are summarized in Table III. The unusually large value of the moment, 5.0 $\mu_B$/Mn, at x=0.1\% requires the Mn to occupy an S site and is not completely due to the Mn alone \cite{Shaughnessy}. The calculated magnetic moments for all cases are given Table II. For the 8-atom case with Mn at the S site, the constricted relaxation due to the small cell size causes the magnetic moment to be 0.03$\mu_B$/Mn less than the value predicted from the ionic model. This indicates there is an interaction between cells. When the size of the cell increases, the calculated magnetic moment agrees with the predicted value for the Mn. 
	
	On the other hand, the magnetic moments at an S site for the Fe doping is only 1.76 $\mu_B$/Fe, much smaller than the predicted 4.0 $\mu_B$/Fe, and are zero for an Fe in 64- and 216-atom cells. As shown in Table I for the Fe at an S site, the bond lengths are 2.32, 2.26 and 2.25 \AA \space for the 8- , 64- and 216-atom cells, respectively. For the 8-atom case, the situation is similar to the case of Mn doping. The restricted contraction and the interaction between the neighboring cells together determine the magnetic moment. For larger supercells, the significant contraction of the bond length renders the volume around the Fe atom too small to align either the four or even the two spin moments and still obey the Pauli exclusion principle. The four electrons oppositely pair their spins in order to coexist in the confined volume. The resultant spin moment for the alloy is zero. 
	
	When a Mn is at an I site, the magnetic moment is larger than 3.00 $\mu_B$/Mn for the 8-atom case while it is 3.00 $\mu_B$/Mn in the other two cases, respectively. As shown in Fig. 4, four of the d-electrons shift their charges into the open regions between bonds formed by the neighboring Si atoms. The three electrons left at the Mn atom align to give 3.00 $\mu_B$/Mn. For the 8-atom case, the cell is small. The four shifted electrons are polarized by the spin moment localized at the Mn through the exchange interaction via the weak overlap of their wave functions. 
	
	With Fe at the I site, the bond lengths expand but not as much as with Mn. For the 8-atom case, the calculated moment is 2.05 $\mu_B$/Fe. The volume around the Fe determined by the bond length of 2.36\AA \space is sufficient to align two spins. The overlap of the wave functions associated with the local spins and the d-states shifting into the interstitial space provide the weak exchange interaction causing the extra 0.05 $\mu_B$/Fe. For the two larger cells, even with greater volumes around the Fe, there is still not enough space to accommodate the alignment of all four spins. One electron flips spin and the resulting moment at the TME is 2.00 $\mu_B$/Fe.
	
	Fully explaining the measured non-integer magnetic moments in Table III requires more sophisticated models that account for randomness and the interaction between dopants atoms, perhaps by considering interactions between supercells or using different models, such as clustering in the alloys instead of one TME per supercell. Our numerical results suggest that in order to get a large magnetic moment for a TME doped Si dilute alloy, it is best to choose a TME having nearly, but less than, half filled d-states.

	\section{Summary}
	We compare the most basic properties of Mn and Fe doped in Si by carrying out first principles calculations based on density functional theory. Alloy models of 8-, 64- and 216-atoms are considered. In each model, two possible dopant sites, the S site and the I site, and two possible dopant species, Fe and Mn, are investigated.		
	
	The relaxation around the dopant at the S site depends on its atomic radius. A contraction is consistently obtained for Fe because its atomic radius is smaller than that of a Si atom. The opposite behavior is determined for doping Mn. At an I site, the surrounding Si atoms move away from either dopant independent of the TME radius because some of the d-states shift their charges toward the open space under the attractive interaction with the Si nuclei. These shifted charges repel the bond charges between the Si atoms and expand the bond near the TME. 
	
		The effect of the contraction due to relaxations can have a detrimental effect on the value of the magnetic moment - diminishing the magnetic moment/TME. The resulting small volume can force all the spins of the metal atom's electrons to pair. A high concentration of either Fe or Mn can constrict the relaxation through dopant-dopant interactions. We make a few suggestions for doping TMEs in Si.
			
	To begin, atoms with close to but less than half filled d-states will be better choices because the valence electron spins can potentially align to yield a large magnetic moment/dopant in the alloy. Secondly, it is better to dope with a TME with a larger atomic radius. As shown in the S site cases, contraction of the lattice around the smaller Fe reduces the available volume and restricts the spin alignment, lowering the moment, while the lattice expansion around the larger Mn at an S site allows for a greater moment.  
	
	Third, near an I site the volume expands independent of atomic radius. The expansion may not, however, be enough to maximize the spin alignment. Fourth, the 8-atom results, most relevant to high doping concentrations, yield unexpected magnetic moments because of coupling between the dopant atoms. Furthermore, the relaxation around TMEs can also exhibit unaniticipated behavior. Therefore, if the doping concentration is over 10\%, one should take the interaction between the dopants into account.  
	
	The first two suggestions are common sense. We confirm them by carrying out explicit numerical calculations. We see that the magnetic properties of the isolated TME are modified by the Si lattice environment. The third and fourth point may not be so intuitively obvious, showing that the expansion at the I site and dopant-dopant interaction between cells affect both lattice relaxations and the resultant magnetic moments at high concentrations.   
	 	
	Finally, we note that doping TMEs at S sites may be possible with low temperature molecular beam epitaxy. Otherwise it is necessary to break bonds for the metal atoms to occupy the S site. Other methods, such as the ion implantation growth scheme, are necessary. There are recent suggestions that co-doping with pnictides can reduce the S site formation energy and increase the magnetic moment per unit cell. \cite{Zhu}

This work was supported by the NSF Grant No. ECCS-0725902. Work at Lawrence Livermore National Laboratory was performed under the auspices of the U.S. Department of Energy under Contract DE-AC52-07NA27344.

\newpage


\newpage

\textbf{Figure Captions}


Fig.1 Left: 8-atom cell with TME at an S site. The TME is the light (yellow online) sphere and the Si atoms are the darker (blue online) spheres. Right: 8-atom cell with TME at an I site.
\newline
\newline


Fig. 2 (a) Diagram showing the d-p hybridization between the Si atom at (0.0,0.0,0.0)a and the TME at (1/4,1/4,1/4)a, where a is the lattice constant of the conventional cell. (b) Schematic diagram showing the crystal field and hybridization effects.
\newline
\newline

Fig. 3 Total charge density of Mn at an S site in the 64-atom model. The section is formed by [110] and [001] axes containing Mn at an S site.
\newline
\newline

Fig. 4 Total charge density of Mn at an I site in the 64-atom model. The section is formed by [110] and [001] axes containing the Mn at an I site.
\newline
\newline
\end{document}